\documentstyle[preprint,aps]{revtex}
\input epsf.tex
\def\DESepsf(#1 width #2){\epsfxsize=#2 \epsfbox{#1}}
\begin{document}
\preprint{\vbox{\hbox{UH-511-915-98}}}
\draft
\title{Electroweak Penguins, Final State Interaction Phases\\
and CP Violation in $B\to K \pi$ Decays}
\author{$^1$N.G. Deshpande, $^{2,3}$Xiao-Gang He, $^{2,4}$Wei-Shu Hou, 
and $^5$Sandip Pakvasa}
\address{$^1$Institute of Theoretical Science, University of Oregon,
Eugene, OR 97403-5203, USA\\
$^2$ Department of Physics, National Taiwan University, 
Taipei, Taiwan 10764, R.O.C.\\
$^3$School of Physics, University of Melbourne,
Parkville, Vic. 3052, Australia\\
$^4$ Physics Department, Brookhaven National Laboratory, 
Upton, NY 11973, USA\\
and\\
$^5$Department of Physics and Astronomy, University of Hawaii,
Honolulu, HI 90822, USA}
\date{September 1998}
\maketitle
\begin{abstract}
The recently observed $B^-\to K^-\pi^0,\ \bar K^0 \pi^-$ 
and $\bar B^0\to K^-\pi^+$ decay modes
appear to have nearly equal branching ratios. 
This suggests that tree and 
electroweak penguins play an important role, 
and inclusion of the latter improves agreement between 
factorization calculation and experimental data.
The value of $\gamma$ in the range of
$90^\circ - 130^\circ$ and $220^\circ - 260^\circ$ is favored, 
while the $\bar B^0\to \bar K^0 \pi^0$ rate is suppressed. 
Direct CP violation for $B\to K \pi$ modes can be 
large if final state interaction phases are large.
\end{abstract}
The CLEO Collaboration has recently made 
the first observation of the decay $B^-\to K^-\pi^0$,
with the branching ratio ($Br$) of
$(1.5\pm 0.4\pm0.3)\times 10^{-5}$ \cite{cleo1}.
They have also remeasured 
$Br(\bar B^0\to K^-\pi^+) = (1.4\pm 0.3\pm 0.2)\times 10^{-5}$ and 
$Br(B^-\to \bar K^0 \pi^-) = (1.4\pm0.5\pm 0.2)\times 10^{-5}$.
Though confirming previous results \cite{cleo2}, 
the central value for the latter has dropped by 40\%.
The three rates now appear remarkably close to one another.
If they are dominated by the strong penguin interaction, then
$Br(B^-\to K^-\pi^0) \sim 1/2\; Br(B^-\to K^- \pi^+)$ is expected.
As errors further improve, if 
$K^-\pi^0 \simeq K^-\pi^+ \simeq \bar K^0 \pi^-$ still persists, 
there would be important implications for 
the interference between the strong penguin, the tree, and especially
the electroweak penguin (EWP) interactions \cite{dh1,fleischer1}, 
final state interaction (FSI) phases
 \cite{fleischer1,hou,isospin,donoghue,soni}
and CP asymmetries ($a_{CP}$) in these decays. 
The isospin related $\bar B^0\to \bar K^0\pi^0$ decay rate 
can also be inferred once the other three are precisely known 
\cite{soni,lipkin}. 
In the following we carry out an analysis in the Standard Model (SM). 
Our conclusions are suggestive and depend on 
the final experimental $Br$s being 
close to the present central values.

We decompose the decay amplitudes according to 
final state isospin\cite{isospin}. 
This enables us to compare strong penguin,  tree
and EWP contributions, 
as well as carry out analyses for FSI phases.
For $B\to K\pi$ decays, 
the $I=1/2$ and $3/2$ amplitudes are generated in SM by 
the $\Delta I = 0$ strong penguin
and the $\Delta I = 0$, $1$ tree and EWP 
effective Hamiltonians $H_0^S$, $H_{0,1}^{T}$ and $H_{0,1}^{W}$. 
Denoting the $I=1/2$ amplitudes generated by $H^{j}_{0,1}$
as $a_1^{j}$, $b_1^{j}$ and 
the $I=3/2$ amplitudes generated by $H_1^{j}$ as $b_3^{j}$,
we write the $A_{K\pi}$ amplitudes as
\begin{eqnarray}
A_{K^-\pi^0} &=& 
 {2\over 3} b_3e^{i\delta_3} + \sqrt{{1\over 3}} 
(a_1+b_1)e^{i\delta_1}, \ \ \ 
A_{\bar K^0\pi^-} =
-{\sqrt{2}\over 3} b_3e^{i\delta_3} + \sqrt{{2\over 3}} 
(a_1+b_1)e^{i\delta_1},\nonumber\\
A_{K^-\pi^+} &=&
{\sqrt{2}\over 3} b_3e^{i\delta_3} + \sqrt{{2\over 3}} 
(a_1-b_1)e^{i\delta_1},\ \ \
A_{\bar K^0\pi^0} = 
{2\over 3} b_3e^{i\delta_3} - \sqrt{{1\over 3}} 
(a_1-b_1)e^{i\delta_1},
\end{eqnarray}
where $a_1 = \sum a_1^j$ with $j$ summed over $T$, $S$ and $W$,
and likewise for $b_i$. 
Since the dominant strong penguin contributes to $a_1$ only,
one expects $A_{K^-\pi^0} \simeq 1/\sqrt{2}\; A_{K^-\pi^+}$
hence $Br(B^- \to K^-\pi^0) \simeq 1/2\; Br(B^0 \to K^-\pi^+)$.
Since the tree amplitude is no more than 20\% of 
the strong penguin amplitude, 
to account for $Br(B^- \to K^-\pi^0) \simeq Br(B^0 \to K^-\pi^+)$,
which violates isospin in the $a_1$-dominance limit,
additional amplitudes such as EWP are called for.
Note that EWP effects do violate isospin.

We have made explicit the elastic $K\pi \to K\pi$ rescattering phases 
$\delta_{1,3}$ in Eq. (1), where only $\delta = \delta_3 - \delta_1$
is physically relevant.
There are additional phases in $a^j_1$ and $b_{1,3}^j$, 
such as the CP violating weak phases in KM matrix elements, 
and absorptive parts due to rescattering between different 
flavored intermediate states with associated KM factors, 
such as charmless and charmed states, 
which cannot be absorbed into $\delta_{1,3}$ \cite{dh1,hou,soni}.
It has been pointed out that FSI phases in B decays
do not necessarily decrease with large $m_B$\cite{donoghue},
and inelastic phases may play a more important role.
However, these phases cannot be calculated, 
while inelastic phases may be subject to cancellations 
when summing over many channels.
Lacking reliable calculations, 
we make the usual approximation of retaining the 
absorptive part from quark level calculation \cite{bander},
ignore long distance inelastic phases, 
and take only the elastic FSI phase $\delta$ to 
model long distance effects \cite{foot}.
Theoretical estimates for $\delta$ have been attempted \cite{donoghue}, 
but we will treat it as a free parameter and 
try to obtain information from data. 

In SM the effective Hamiltonian relevant for Eq. (1) is \cite{buras,dh2}
\begin{eqnarray}
H_{eff} &=& {G_F\over \sqrt{2}}[V_{ub}V^*_{us}(c_1O_{1} + c_2 O_{2}) -
\sum_{i=3}^{10}(V_{ub}V^*_{us} c_i^u
+V_{cb}V^*_{cs} c_i^c +V_{tb}V^*_{ts} c_i^t) O_i] +H.C.\;,
\end{eqnarray}
where the superscripts $u,\;c,\;t$ are for internal quarks.
The operators $O_i$ and the Wilson coefficients (WC) $c^j_i$ are
given explicitly in Ref.\cite{dh2}.
To obtain exclusive decay amplitudes, 
one has to evaluate relevant hadronic matrix elements.  
We use the factorization approximation to estimate the magnitudes, 
then insert the FSI phases $\delta_{1,3}$ as in Eq. (1).
We find
\begin{eqnarray}
&&a_1^T = i{\sqrt{3}\over 4}\, V_{ub}V_{us}^*\; r 
\left[{c_1\over N}+c_2\right],\nonumber\\
&&b_1^T = i {1\over 2\sqrt{3}}\, V_{ub}V_{us}^*\; r 
\left[-{1\over 2}\left({c_1\over N}+c_2\right) 
+\left(c_1+{c_2\over N}\right) X\right],\nonumber\\
&&b_3^T = i{1\over 2}\, V_{ub}V_{us}^*\; r 
\left[\left({c_1\over N}+c_2\right) + \left(c_1+{c_2\over N}\right) X\right],
\nonumber\\
&&a_1^S = -i{\sqrt{3}\over 2}\, V_{ib}V_{is}^*\; r
\left[{c^i_3\over N} + c_4^i + \left({c_5^i\over N}+c_6^i\right)Y\right],
\ \ \ \ \ \ b_1^S = b_3^S = 0\nonumber\\
&&a_1^W = -i{\sqrt{3}\over 8}\, V_{ib}V_{is}^*\; r
\left[\left({c_7^i\over N}+c_8^i\right)Y + {c^i_9\over N} + c_{10}^i\right],
\nonumber\\
&&b_1^W = i{\sqrt{3}\over 4}\, V_{ib}V_{is}^*\; r
\left[{1\over 2}\left(\left({c_7^i\over N}+c_8^i\right)Y 
+ {c^i_9\over N} + c_{10}^i\right) 
+\left(c_7^i+{c_8^i\over N} - c_9^i-{c_{10}^i\over N}\right)X\right],
\nonumber\\
&&b_3^W =- i{3\over 4}\, V_{ib}V_{is}^*\; r
\left[\left(\left({c_7^i\over N}+c_8^i\right)Y 
+ {c^i_9\over N} + c_{10}^i\right) 
-\left(c_7^i+{c_8^i\over N} - c_9^i-{c_{10}^i\over N}\right)X\right],
\end{eqnarray}
where $r = G_F\, f_K F^{B\pi}_0(m^2_K) (m_B^2-m_\pi^2)$,
$X= (f_\pi/f_K)(F^{BK}_0(m^2_\pi)/F^{B\pi}_0(m^2_K)) 
(m_B^2-m^2_K)/(m_B^2-m_\pi^2)$, 
$Y = 2m^2_K/[(m_s+ m_q)(m_b-m_q)]$ with $q=u,\ d$ for 
$\pi^{\pm,0}$ final states, respectively,
and $N$ is the number of effective colors.
We have neglected small annihilation contributions. 
Setting $\delta=0$, we obtain the usual factorization 
result where the tree contribution to $A_{\bar K^0 \pi^-}$ vanish.
Note that $A_{K^-\pi^0}^T \simeq 1/\sqrt{2}\; A_{K^-\pi^+}^T$ 
as well for $\delta = 0$ up to color suppressed corrections.
 
For our numerical calculations we use \cite{ali} 
$f_\pi = 133$ MeV, $f_K = 158$ MeV, $F^{B\pi}_0(0) = 0.36$, 
$F^{BK}_0(0)=0.41$, assume monopole dependence for $F_0(k^2)$, 
and use the $c_i^j$ values obtained in Ref. \cite{dh2}.
The WC's are scheme dependent, which should be compensated by
hadronic matrix elements evaluated in the same scheme. 
Unfortunately, there is no reliable way to 
calcualte the hadronic matrix elements at present. 
The uncertainties are large for absolute $Br$s, 
mainly from form factors, 
while we have little control of non-factorizable effects. 
But rather than aiming at precise predictions, 
we wish to demonstrate that the measured $Br$s 
can be accommodated within uncertainties. 
The relative strength of strong penguin, tree and EWP contributions, 
hence the ratios of branching ratios, 
are much less sensitive to the input values.
The same can be said about $a_{CP}$s except 
for uncertainties in FSI \cite{foot}.

There is an additional uncertainty in $a_{CP}$
from the value of $q^2$ to take \cite{hou}.
Care also has to be taken to 
include absorptive parts from the gluon propagator.
The quark level absorptive parts due to $\bar cc$ rescattering
depend on the $q^2$ of the virtual gluon, 
which is not well defined for exclusive processes.
Although the $Br$s are not very sensitive to the specific value of $q^2$, 
the $a_{CP}$s are sensitive to $q^2$ when $\delta$ is small. 
Two configurations need to be distinguished. 
When the pion comes off from the $q'\bar q'$ current in the penguin, 
$q^2$ should take the value of $m_\pi^2$. 
In case of Fierz transformed operators, 
the kaon contains the $s$ quark and the  $\bar q'$ quark.
We assume that the two light quarks share the kaon momentum equally, 
then $q^2$ is approximately given by $m_b^2/2$,
which is favorable for large $a_{CP}$s when the FSI phase is small. 
The $a_{CP}$s obtained this way
should be viewed as an upper limit for $\delta =0$. 
But for large values of $\delta$, 
the choice of $q^2$ becomes unimportant.

Our results are given in Figs. 1 to 3, 
where $Br$s are averaged over $B$ and $\bar B$ decays.
We shall first explore the case without EWP contributions, 
as naively they are suppressed by $\alpha \sim 1/137$.
Although the value of $\gamma$ is still not well determined
and is a topic we shall study, we use $\gamma = 64^\circ$ 
obtained in Ref. \cite{gamma} for illustration.
In Fig. 1(a) we show the dependence of $Br$s on FSI phase $\delta$.
We see that 
$ K^- \pi^0/K^- \pi^+ \sim \bar K^0 \pi^0/\bar K^0 \pi^- \sim 1/2$ 
for all $\delta$, which clearly indicates strong penguin dominance. 
This is a general feature without EWPs 
which was pointed out in Ref. \cite{hhy}. 
For large $\delta$, the $\bar K^0 \pi^-$ and $K^- \pi^+$ rates
approach each other.

As pointed out some time ago, electroweak penguins 
are important in some $B\to K \pi$ decay modes \cite{dh1}.  
Adding the EWP effect, the results are given in Fig. 1(b).
The impact is quite visible. 
Not only the $\delta$ dependence is different, but the most salient is
that the $K^-\pi^0$ and $K^-\pi^+$ rates become much closer to each other.
However, they lie considerably below the  $B^-\to \bar K^0 \pi^-$ mode,
and the splitting reaches maximum at $\delta = 180^\circ$. 
Clearly the data prefer smaller values of $\delta$, 
but larger values are not ruled out within the errors at present.
For small $\delta$, the $\bar B^0\to \bar K^0\pi^0$ mode
is about two to three times smaller than the other three, 
which agrees with some other estimates\cite{soni,lipkin}.

It is clear that the branching ratios in Fig. 1 cannot give 
$K^-\pi^0 \simeq K^-\pi^+ \simeq \bar K^0 \pi^-$. 
Can they be brought closer to one another? 
The $Br$s depend strongly on $\gamma$ through the tree amplitude,
which offers an extra handle. 
We give the result without EWP in Fig. 2(a) for $\delta = 0$.
Both $K^-\pi^+$ and $K^-\pi^0$ modes exhibit similar
strong dependence on $\gamma$.
However, they remain widely separated, and 
$K^-\pi^0 \simeq K^-\pi^+ \simeq \bar K^0 \pi^-$ 
still cannot be realized.
Turning on EWP contributions, 
the $\gamma$ dependence is given in Fig. 2(b). 
Remarkably, 
it combines both the nice features of Figs. 1(b) and 2(a): 
$K^-\pi^0$ and $K^-\pi^+$ rates are close to each other in value,
and they exhibit similar strong dependence on $\gamma$.
To have the three branching ratios
within one standard deviation of the experimental central values,
$\gamma$ is preferred to be in 
the ranges of $90^\circ - 130^\circ$ and $220^\circ - 260^\circ$,
which are in different quadrants from $\gamma=64^\circ$ \cite{gamma}.
The branching ratios are large enough such that
there is no need to scale up the form factors,
but $K^- \pi^0$ is never larger than $K^- \pi^+$.
The corresponding $\bar K^0 \pi^0$ rate
is typically three times smaller.
Note that the method proposed to constrain 
$\gamma$ from $K^-\pi^+/\bar K^0 \pi^-$ \cite{fleischer}
is no longer useful since the ratio is now almost one.
However, the near equality of the three observed $Br$s 
favors \cite{nr} $\gamma$ in the range of
$90^\circ-130^\circ$ and $220^\circ - 260^\circ$
when EWP effects are included.

A direct way of measuring the strength of the electroweak penguin is to 
observe modes such as $B\to K^{(*)}\ell^+\ell^-$ \cite{hws} 
or $\bar B_s\to \pi (\eta, \phi)$\cite{dh3},
but these rates are small and not yet measured. 
The study of $B\to K\pi$ modes thus provide a 
more practical method of probing EWP effects through interference,
where the presently observed near equality of
$K^-\pi^0$, $K^-\pi^+$ and $\bar K^0\pi^-$ modes
seems to offer a unique window. 
The dominant effect comes from $c_9$ and $c_{10}$,
which arise from the $Z$ penguin \cite{hws},
and the effective strength is comparable to tree.
Furthermore, 
these terms violate isospin and come in the right sign and strength
to bring the $K^-\pi^0$ and $K^-\pi^+$ rates close together.

Direct CP violating partial rate asymmetries are of great interest. 
In Fig. 3(a) we show the $\gamma$ dependence with $\delta = 0$,
where we now always include EWPs. 
The $a_{CP}$s can be as large as 13\% for $K^-\pi^+$ 
and 8\% for $K^- \pi^0$, 
but they are small for both $\bar K^0\pi^-$ 
and $\bar K^0\pi^0$ 
because the tree contribution to the former is zero and the latter 
is color suppressed. 
In Fig. 3(b) we show the dependence of the $a_{CP}$s on $\delta$
for $\gamma = 120^\circ$, 
a typical value in the preferred range suggested by $K\pi$ data.
Results for $\gamma = 64^\circ$ and $240^\circ$
are qualitatively similar.
For $\delta$ around $10^\circ$-$20^\circ$ as 
suggested by some theoretical calculations \cite{donoghue},
the $a_{CP}$s vary considerably, with
$\bar K^0 \pi^-$ and $\bar K^0 \pi^0$ gaining while 
$K^-\pi^+$ and $ K^- \pi^0$ drop.
If the phase $\delta$ turns out to be large,
$a_{CP}$s can be as large as -20\% to 40\%,
with charged and neutral kaon modes typically having opposite sign.
However, large $\delta$ values would again split
the $\bar K^0 \pi^-$ mode upwards from the other two observed modes,
although $\vert \delta\vert < 60^\circ$ is still allowed.

To see how non-factorizable effects may affect our results,
we vary the effective number of colors $N$. 
We find that a smaller $N$ lowers 
the $Br$s but brings them closer to each other.
Since the value of $m_s$ is not well known, we note that
a smaller $m_s$ enhances the strong penguin contribution and
tends to enforce $K^-\pi^0 \sim 1/2\; K^-\pi^+$ \cite{hhy}. 
At present $1/N$ between 1/2 to zero 
and $m_s$ between 100 MeV to 200 MeV are allowed by data.
Improvements on experimental $Br$s will 
further restrict the allowed ranges for the parameters involved. 

The $a_{CP}$s discussed above depend strongly on the FSI phase $\delta$. 
If the rate differences for all the $B\to K \pi$ modes are measured, 
one can observe CP violation involving only $I=1/2$ amplitudes 
which is free from $\delta$.
This is in contrast with the kaon system where, 
in order to have rate asymmetry,
there must be at least two isospin amplitudes \cite{dhp}.
Let us define
\begin{eqnarray}
\Delta & = & \Delta_{K^-\pi^0} + \Delta_{\bar K^0\pi^-}
            - \Delta_{K^-\pi^+} - \Delta_{\bar K^0 \pi^0}\nonumber\\
     & \sim & |a_1+b_1|^2 - |\bar a_1 + \bar b_1|^2
            - |a_1-b_1|^2 + |\bar a_1-\bar b_1|^2,
\end{eqnarray}
where $\Delta_{ij}$ is the rate difference between $B$ and $\bar B$ decays, 
and barred amplitudes refer to anti-particles. 
Normalizing to 
$\Gamma(\bar B^0 \to K^- \pi^+$), 
we find that the quark level calculation gives
$\Delta/ \Gamma(\bar B^0\to K^- \pi^+)$ around $-5\%$ (-3\%) 
for $\gamma = 64^\circ\ (120^\circ)$, 
which can be tested at B factories. 
This measurement also serves as a test for any additional relative phase 
between $a_1$ and $b_1$ as could arise from long distance inelastic FSI. 
For example, putting an additional CP conserving phase \cite{foot} 
of $20^\circ,\ 70^\circ, -20^\circ, -70^\circ$ in $a_1$, 
the value $\Delta/\Gamma(\bar B^0\to K^- \pi^+)$ is about $-12\%$, $-19\%,\;
4\%,\;18\%$ 
($-8\%$,$-18\%,\;2.4\%,\;14\%$)
for $\gamma = 64^\circ\ (120^\circ)$.

In conclusion, the present data on $B\to K\pi$ decay modes 
suggest that the electroweak penguins are important.
Without them the $\bar B\to \bar K^0 \pi^-$ or $K^- \pi^+$ rates would
be considerably higher than that of $K^- \pi^0$. 
This makes the study of $B\to K\pi$ modes a more practical way to probe the
hadronic electroweak penguin effects than $B_s\to \pi(\eta, \phi)$.
The branching ratio for $\bar B^0\to \bar K^0\pi^0$
is predicted to be a factor of two to three smaller 
if the other three decay rates are approximately equal. 
The present experimental data still allow a FSI phase $\delta$
up to $60^\circ$. 
CP violation in $\bar B\to K^- \pi^0$, $\bar K^0 \pi^-$, $K^-\pi^+$ 
and $\bar K^0 \pi^0$ modes can be as large as 
30\%, 15\%, 25\% and 40\%, respectively,
with characteristic sign correlations.

We thank T.E. Browder and M. Suzuki 
for useful discussions and a careful reading of 
the manuscript.
This work was supported in part by 
US Department of Energy Grant No. DE-FG06-85ER40224,  DE-AN03-76SF00235, 
by ROC National Science Council Grant No. NSC 88-2112-M-002-033, 
and by Australian Research Council.

\begin{figure}[htb]
\centerline{ \DESepsf(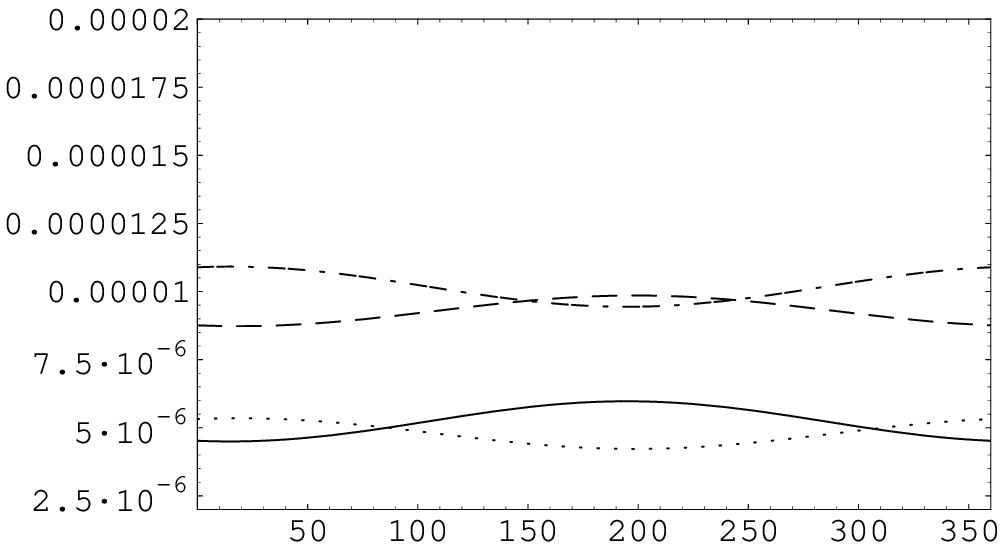 width 8 cm)
             \DESepsf(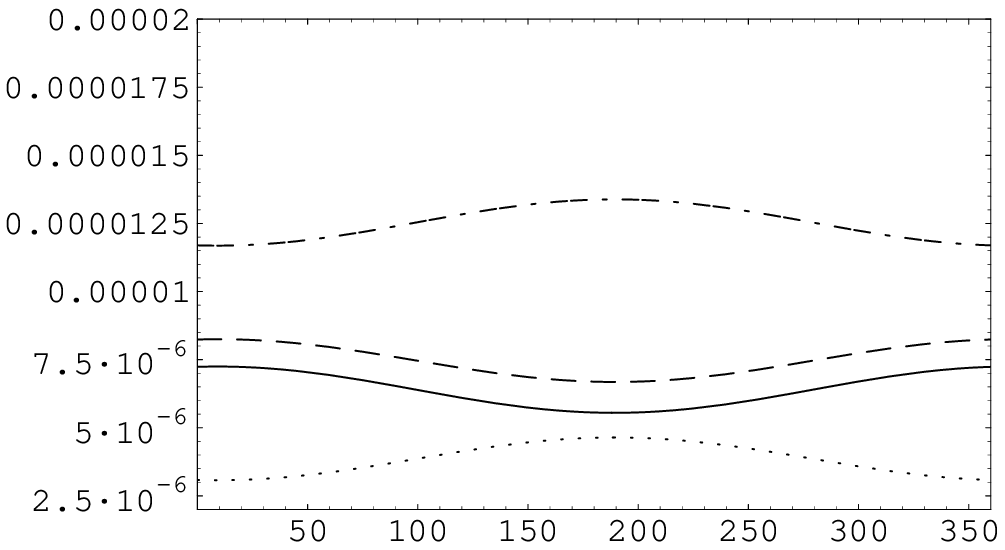 width 8 cm) }
\smallskip
\caption {
$Br(B\to K\pi)$ vs. $\delta$ for $\gamma = 64^\circ$ 
(a) without or (b) with electroweak penguin contributions.
In all the figures we use $N= 3$ and $m_s = 200$ MeV. 
Solid, dot-dashed, dashed and dotted lines are for 
$B^-\to K^-\pi^0,\ \bar K^0 \pi^-$ and 
$\bar B^0 \to K^-\pi^+,\ \bar K^0\pi^0$,
respectively.
}
\label{gamma1}
\end{figure}

\begin{figure}[htb]
\centerline{ \DESepsf(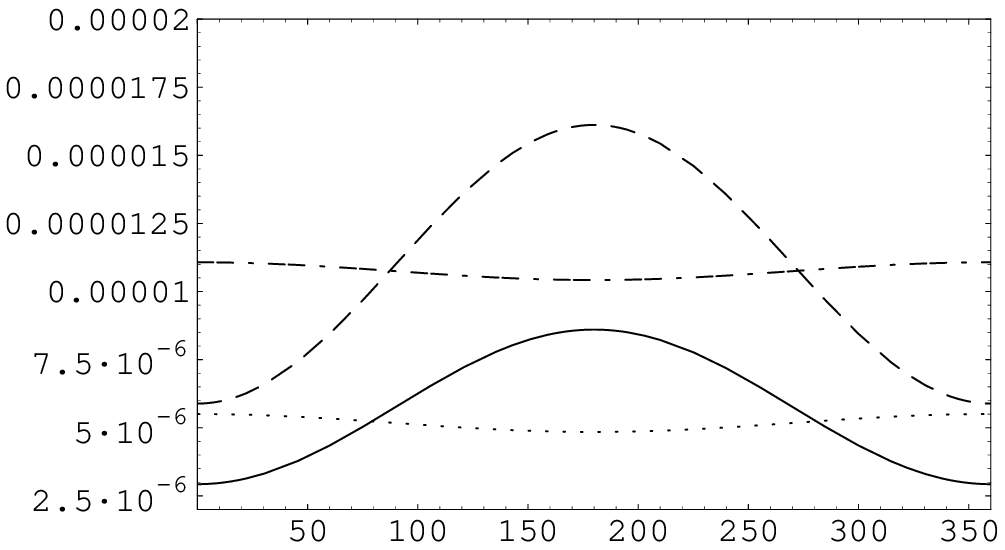 width 8 cm)
             \DESepsf(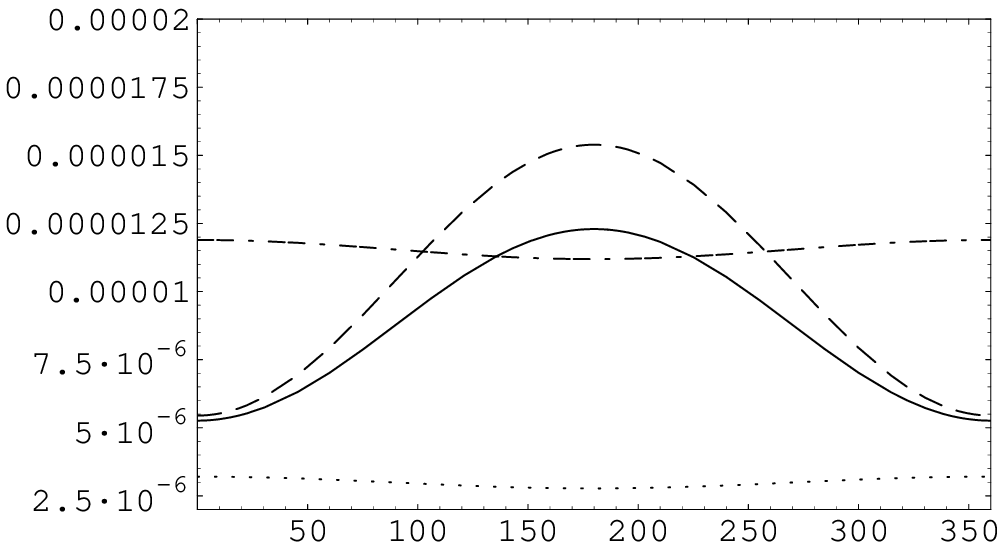 width 8 cm) }
\smallskip
\caption {
As in Fig. 1 but vs. $\gamma$ for $\delta = 0$.
}
\label{gamma2}
\end{figure}

\begin{figure}[htb]
\centerline{ \DESepsf(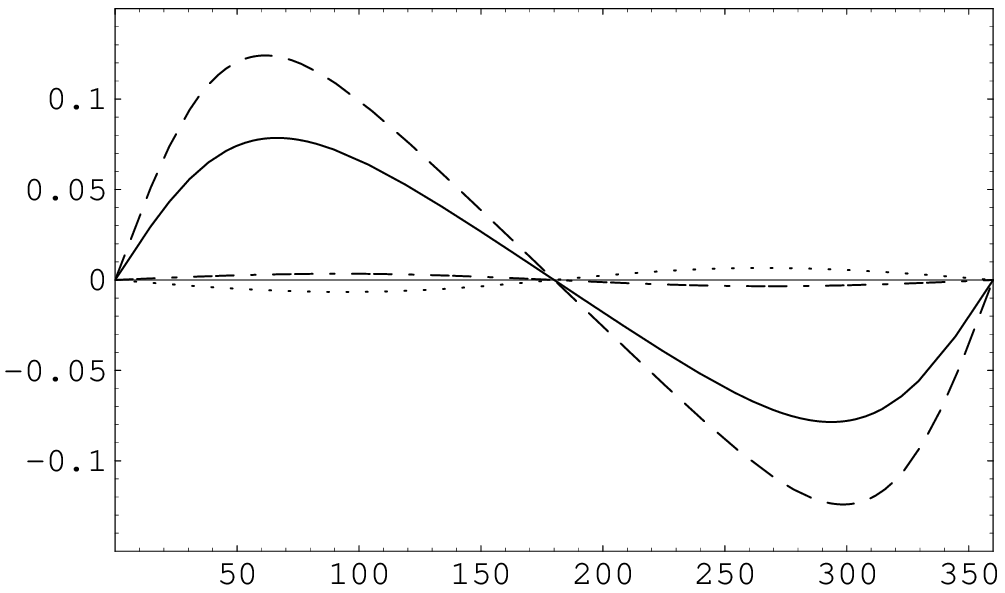 width 8 cm)
             \DESepsf(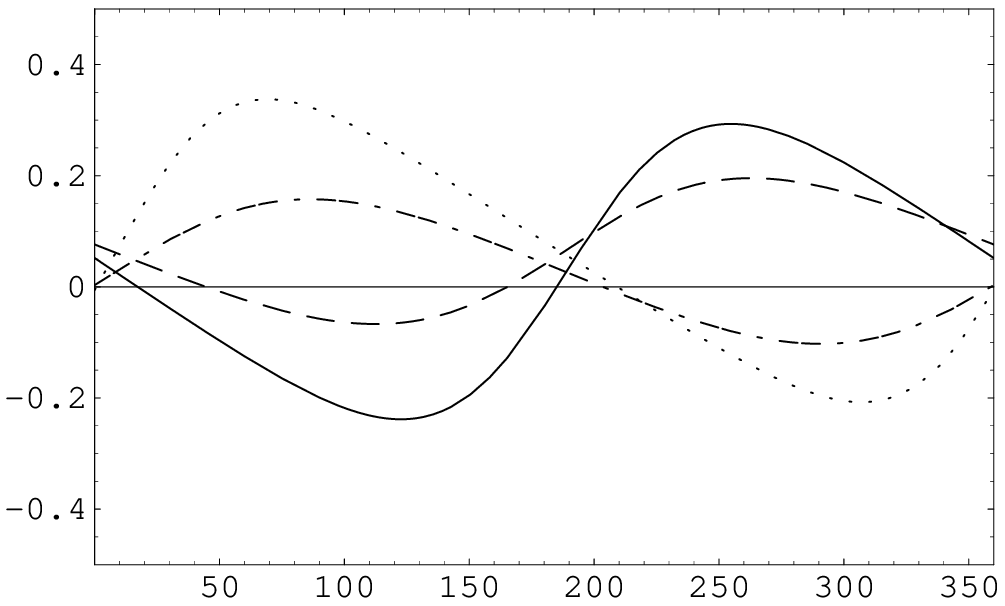 width 7.9 cm) }
\smallskip
\caption {$a_{CP}(B\to K \pi)$ with EWP contributions 
(a) vs. $\gamma$ for $\delta = 0$ and
(b) vs. $\delta$ for $\gamma = 120^\circ$. 
}
\label{gamma5}
\end{figure}


\begin{references}
\bibitem{cleo1} M. Artuso et al. (CLEO Collaboration), CLEO CONF 98-20.

\bibitem{cleo2} R. Godang et al. (CLEO Collaboration), 
Phys. Rev. Lett. {\bf 80}, 3456 (1998).

\bibitem{dh1} N.G. Deshpande and X.G. He, 
Phys. Rev. Lett. {\bf 74}, 26 (1995); 4099(E) (1995). 

\bibitem{fleischer1} A. Buras and R. Fleischer, Phys. Lett. 
{\bf B365}, 390(1996); R. Fleischer, ibid, 399 (1996).

\bibitem{hou} J.-M. G\' erard and W.S. Hou, 
Phys. Rev. Lett. {\bf 62}, 855 (1989); Phys. Rev. {\bf D43}, 2909 (1991).

\bibitem{isospin} Y. Nir and H. Quinn, Phys. Rev. Lett. {\bf 67}, 541(1991);
M. Gronau, Phys. Lett. {\bf B265}, 389(1991).
We use the notation of J.-M. G\' erard and J. Weyers, e-print hep-ph/9711469. 

\bibitem{donoghue} J. Donoghue et al., Phys. Rev. Lett. {\bf 77}, 2178(1996).
For further discussion, see 
M. Neubert, Phys. lett. {\bf B424}, 152 (1998);
A. Falk et al., Phys. Rev. {\bf D57}, 4290 (1998);
D. Del\' epine et al., Phys. Lett. {\bf B429}, 106 (1998);
and M. Suzuki, e-print hep-ph/9807414 and hep-ph/9808303.

\bibitem{soni} D. Atwood and A. Soni, Phys. Rev. {\bf D58}, 036005 (1998).

\bibitem{lipkin} H. Lipkin, e-print hep-ph/9809328.

\bibitem{bander} M. Bander, D. Silverman and A. Soni, 
Phys. Rev. Lett. {\bf 43}, 242 (1979).

\bibitem{foot} We can model long distance inelastic FSI phases by 
putting in an additional relative phase between $a_1$ and $b_1$,
as we shall illustrate. 
However, due to different KM factors in $a_1$ and $b_{1}$,
one cannot isolate a pure strong phase as in the elastic case.
At any rate, predictive power is lost.

\bibitem{buras} M. Lautenbacher and P. Weisz, Nucl. Phys. {\bf B400}, 37 (1993); 
A. Buras, M. Jamin and M. Lautenbacher, ibid. {\bf B400}, 75 (1993);
M. Ciuchini et al., 
ibid. {\bf B415}, 403  (1994).

\bibitem{dh2} N. G. Deshpande and X.G. He, 
Phys. Lett. {\bf B336}, 471 (1994).

\bibitem{ali} A. Ali, G. Kramer and C.-D. Lu, e-print hep-ph/9804363.


\bibitem{gamma} See, for example, 
F. Parodi, P. Roudeau and A. Stocchi, e-print hep-ph/9802289.

\bibitem{hhy} X.G. He, W.S. Hou and K.C. Yang, e-print hep-ph/9809282,
to appear in Phys. Rev. Lett.

\bibitem{fleischer} R. Fleischer and T. Mannel, 
Phys. Rev. {\bf D57}, 2752 (1998).

\bibitem{nr} See also M. Neubert and J.L. Rosner, 
e-print hep-ph/9808493, to appear in Phys. Lett. {\bf B}.

\bibitem{hws} W.S. Hou, R.S. Willey and A. Soni, 
Phys. Rev. Lett. {\bf 58}, 1608 (1987).

\bibitem{dh3} N.G. Deshpande, X.G. He and J. Trampetic,
Phys. Lett. {\bf B345}, 547 (1995).

\bibitem{dhp} N.G. Deshpande, X.G. He and S. Pakvasa, e-print hep-ph/9606259.
\end{references}
\end{document}